\documentclass[12pt]{article} 

\usepackage[sectionbib]{natbib}
\usepackage{array,epsfig,fancyheadings,rotating}
\usepackage[]{hyperref}  
\usepackage{sectsty, secdot}
\sectionfont{\fontsize{12}{14pt plus.8pt minus .6pt}\selectfont}
\renewcommand{\theequation}{\thesection\arabic{equation}}
\subsectionfont{\fontsize{12}{14pt plus.8pt minus .6pt}\selectfont}

\usepackage{lscape}
\usepackage{amsmath}
\usepackage{amssymb}
\usepackage{amsfonts}
\usepackage{multirow}
\usepackage{amsthm}
\usepackage[boxed]{algorithm2e}
\usepackage{subfigure}
\usepackage{threeparttable}
\usepackage{caption}
\usepackage{natbib}
\usepackage{multirow}
\usepackage{setspace}
\usepackage{booktabs}
\usepackage{lineno}

\newtheorem{theorem}{Theorem}
\newtheorem{lemma}{Lemma}

\theoremstyle{definition}

\hypersetup{
colorlinks=true,
linkcolor=blue,
citecolor=blue,
urlcolor=blue}

\def\beq{\begin{equation}}
\def\eeq{\end{equation}}
\def\beqr{\begin{eqnarray}}
\def\eeqr{\end{eqnarray}}
\def\beqrs{\begin{eqnarray*}}
\def\eeqrs{\end{eqnarray*}}
\def\bet{\begin{theorem}}
\def\eet{\end{theorem}}
\def\bel{\begin{lemma}}
\def\eel{\end{lemma}}
\def\bg{\begin{figure}[tbph]\begin{center}}
\def\eg{\end{center}\end{figure}}

\def\bc{\begin{center}}
\def\ec{\end{center}}

\def\1{\mbox{\boldmath $1$}}

\def\mS{\mathcal S}

\def\mcO{\mathcal O}

\def\thetaSOS{\widehat{\theta}_{_{SOS}}}
\def\thetaJDS{\widehat{\theta}_{_{JDS}}}

\def\var{\operatorname{var}}

\def\wh{\widehat}

\def\DeltaSOSone{\wh \Delta^{(1)}_{_{SOS}}}
\def\DeltaSOStwo{\wh \Delta^{(2)}_{_{SOS}}}
\def\DeltaJDSone{\wh \Delta^{(1)}_{_{JDS}}}
\def\DeltaJDStwo{\wh \Delta^{(2)}_{_{JDS}}}

\begin{document}


\renewcommand{\baselinestretch}{2}

\markright{ \hbox{\footnotesize\rm  
}\hfill\\[-13pt]
\hbox{\footnotesize\rm
}\hfill }

\markboth{\hfill{\footnotesize\rm Shuyuan Wu, Xuening Zhu, and Hansheng Wang} \hfill}
{\hfill {\footnotesize\rm APPENDIX} \hfill}

\renewcommand{\thefootnote}{}
$\ $\par


\fontsize{12}{14pt plus.8pt minus .6pt}\selectfont \vspace{0.8pc}
\centerline{\large\bf Subsampling and Jackknifing: A Practically Convenient }
\centerline{\large\bf Solution  for Large Data Analysis with Limited }
\centerline{\large\bf Computational Resources }
\vspace{.4cm}
\centerline{ Shuyuan Wu$^1$, Xuening Zhu$^{2}$, and Hansheng Wang$^1$}
\vspace{.4cm}
\centerline {\it $^1$Guanghua School of Management, Peking University, Beijing, China }
\centerline{\it $^2$School of Data Science, Fudan University, Shanghai, China}
\vspace{.55cm} \fontsize{9}{11.5pt plus.8pt minus.6pt}\selectfont

\begin{singlespace}
\begin{footnotetext}[1] {
\noindent
Corresponding author: Xuening Zhu, School of Data Science, Fudan University, Shanghai, China. E-mail:
\href{xueningzhu@fudan.edu.cn}{xueningzhu@fudan.edu.cn}.
}
\end{footnotetext}
\end{singlespace}

\begin{quotation}
\noindent {\it Abstract:}
Modern statistical analysis often encounters datasets with large sizes. For these datasets, conventional estimation methods can hardly be used immediately because practitioners often suffer from limited computational resources. In most cases, they do not have powerful computational resources (e.g., Hadoop or Spark). How to practically analyze large datasets with limited computational resources then becomes a problem of great importance. To solve this problem, we propose here a novel subsampling-based method with jackknifing. The key idea is to treat the whole sample data as if they were the population. Then, multiple subsamples with greatly reduced sizes are obtained by the method of simple random sampling with replacement.  It is remarkable that we do not recommend sampling methods without replacement because this would incur a significant cost for data processing on the hard drive. Such cost does not exist if the data are processed in memory. Because subsampled data have relatively small sizes, they can be comfortably read into computer memory as a whole and then processed easily. Based on subsampled datasets, jackknife-debiased estimators can be obtained for the target parameter. The resulting estimators are statistically consistent, with an extremely small bias. Finally, the jackknife-debiased estimators from different subsamples are averaged together to form the final estimator. We theoretically show that the final estimator is consistent and asymptotically normal. Its asymptotic statistical efficiency can be as good as that of the whole sample estimator under very mild conditions. The proposed method is simple enough to be easily implemented on most practical computer systems and thus should have very wide applicability.

\vspace{9pt}
\par
\end{quotation}\par

\def\thefigure{\arabic{figure}}
\def\thetable{\arabic{table}}

\renewcommand{\theequation}{\thesection.\arabic{equation}}

\fontsize{12}{14pt plus.8pt minus .6pt}\selectfont

\section{ INTRODUCTION}
\noindent

Modern statistical analysis often encounters datasets with large sizes. Meanwhile, most researchers possess very limited computational resources. In most cases, they do not have a powerful computation system, such as a distributed computation system like Hadoop or Spark. As a consequence, they must rely on their handy computational resources (e.g., a personal computer) for large data analysis. Thus, how to practically analyze large datasets with limited computational resources becomes a problem of great importance.

To solve this problem, various subsampling methods have been proposed \citep{mahoney2011randomized,drineas2011faster,ma2015statistical,wang2018optimal,wang2019more,Aimingyao2020,ma2020asymptotic}. The key idea of most existing methods is to design a novel sampling strategy so that excellent statistical efficiency can be achieved with small sample sizes. For example, \citet{ma2015statistical} developed a novel method to select the optimal subsample according to leverage scores. \citet{wang2018optimal} studied a similar problem and proposed an $A$-optimality criterion. \citet{Aimingyao2020} developed an optimal Poisson subsampling approach.  \citet{ma2020asymptotic} derived the asymptotic distribution of the sampling estimator based on the linear regression. Despite their usefulness, these pioneer methods suffer from two limitations. First, specific sampling strategies must be carefully designed for different analysis purposes. Second, they are not computationally inexpensive. A significant computation cost is required for practical implementation. In most cases, the sampling cost should be at least $O(N),$ where $N$ represents the whole sample size.

To overcome these challenges, here, we aim to develop a novel method with the following unique features. First, our method is simple enough to be easily implemented on most practical computer systems. We argue that the simplicity is particularly relevant and important because simplicity implies wider applicability. Second, due to jackknifing, our estimators lead to significant bias reduction compared with other methods. As a result, the same asymptotic efficiency can be achieved with a much reduced subsample size as long as the number of subsamples is large enough.  Moreover, our method supports fully automatic and unified inference. For most real applications, valid statistical inferences (e.g. confidence interval) are inevitably needed. However, the analytical formula for the asymptotic distribution of the estimator could be too complicated to be analytically derived. Our proposal is automatic in the sense that the standard errors of various statistics can be automatically computed without referring to the analytical formula of their asymptotic distributions. In addition, our proposal is unified in the sense that it can be readily applied to many different statistics.

Specifically, we develop here a subsampling method with jackknifing. To implement our method, multiple subsamples are obtained by simple random sampling with replacement.
For each subsampled dataset, a jackknife-debiased estimator is computed for the parameter of interest. Subsequently, these jackknife-debiased estimators are further averaged. This leads to the final estimator. We show theoretically that the resulting estimator is consistent and asymptotically normal. Its statistical efficiency can be asymptotically as good as the whole sample estimator under very mild conditions. This useful property remains valid even if the subsample size is very small. The desirable property is mainly attributed to jackknifing. As a byproduct, a jackknife estimator for the standard error of the proposed estimator can be obtained. This enables automatic statistical inference. For practical implementation, a GPU-based algorithm is developed. Empirical experiments suggest that it is extremely computationally efficient. Extensive numerical studies are presented to demonstrate the finite sample performance.

Despite the usefulness, the proposed method also suffers from obvious limitations. The main limitation is that it is computationally less efficient as compared to the one-pass-full-sample-mean estimator computed by the distributed approaches \citep{suresh2017distributed}.
However, our proposal carries its unique value because it could be a practically more convenient alternative under the following two important situations.
The first situation is that the whole sample size $N$ is extremely large.
In this case, a significant amount of clock-time cost has to be paid for the whole sample computation (e.g., computing the one-pass-full-sample-mean).
This is particularly true if no powerful distributed computation system is available.  However, for most practical data analysis, the practical demand for estimation precision is limited.
On the contrary, the budget for time spending as measured by clock-time cost is extremely
valuable.
Then, it might be more appealing to sacrifice the statistical efficiency to some extent to trade for less clock-time cost.
Accordingly, we do NOT expect our method to be implemented
with a very large subsample size $n$ and a very large number of subsamples $K$. Instead, they should be implemented with reasonably large $n$ and $K$, as long as the desired
statistical precision can be achieved.


The second situation is that automatic statistical inferences are required as we previously mentioned.
In this case, if the one-pass-full-sample-mean  is used, then the analytical formula for the asymptotic distribution of the estimator has to be manually derived.
It is then preferable to have an automatic and unified solution for statistical inference. This is another case where our method could be a practically more convenient solution.

The rest of the paper is organized as follows. Section 2 develops the proposed estimators and their asymptotic properties. The numerical studies are presented in Section 3, including the GPU-based algorithm, simulation experiments, and real dataset analysis. Finally, the article concludes with a brief discussion in Section 4. All technical details are delegated to the Appendix.

\section{THE METHODOLOGY}
\subsection{Model and Notations}
\noindent

Let $X_i$ be an independent random variable observed from the $i$th subject, where $1 \leq i \leq N$ and $N$ is the whole sample size. Let $\mathbb{S} = \left\{1,\dots,N\right\}$ be the index set of the whole sample. Let $\mu$ be one particular moment about $X_i$. For simplicity, we can assume $\mu = E(X_i)$ to be a scalar and that $X_i$ has finite moments. The theory to be presented hereafter can be easily extended to a more general situation with multivariate moments and $M$ estimators. Let $\theta = g(\mu)$ be the parameter of interest, where $g(\cdot)$ is a known nonlinear function. We assume that $g(\cdot)$ is sufficiently smooth.
To estimate $\theta$, one can use a sample moment estimator $\widehat{\theta}$ = $g(\widehat{\mu})$, where $\widehat{\mu} = N^{-1} \sum_{i \in \mathbb{S}} X_i$.

For convenience, we refer to $\widehat{\theta}$ as the whole sample (WS) estimator to emphasize the fact that this is an estimator computed based on the whole sample. The merit of the WS estimator $\widehat{\theta}$ is that it offers excellent statistical efficiency. However, it could be difficult to compute if the whole sample size $N$ is too large. This is particularly true if researchers are given very limited computational resources. Accordingly, one must consider other estimation methods that are more computationally feasible. In this regard, here, we study one particular type of subsampling method \citep{mahoney2011randomized,drineas2011faster,ma2015statistical,wang2018optimal,wang2019more,Aimingyao2020,ma2020asymptotic} as an excellent and practical solution.

Let $n$ be the subsample size, which is typically much smaller than $N$. Let $K$ be the number of subsamples. Write $\mathcal{S}_k = \{i_1^{(k)},\dots,i_n^{(k)}\} \subset \mathbb{S} $ as the $k$th subsample set, where $i^{(k)}_m$s (for any $1\leq m\leq n$, $1\leq k\leq K$) are generated independently from $\mathbb{S}$ by the method of simple random sampling with replacement. In other words, conditional on $\mathbb{S}$, $i^{(k)}_m$s are independently and identically distributed with probability $\mathrm{P}(i^{(k)}_m = j )= N^{-1}$ for any $ j \in \mathbb{S}.$ Accordingly, a moment estimator based on $\mathcal{S}_k$ can be computed as $\widehat \theta^{(k)} = g \left(\widehat{\mu}^{(k)}\right)$, where $ \widehat{\mu}^{(k)} = n^{-1} \sum_{i\in\mathcal{S}_k} X_i.$ One can then combine these subsample estimators together to form a more accurate one as $\thetaSOS = K^{-1} \sum_{k=1}^K \widehat \theta^{(k)}.$ This is then referred to as a subsample one-shot (SOS) estimator. It is similar to the so-called one-shot estimator developed for distributed systems \citep{mcdonald2009efficient,Zinkevich2011Parallelized, zhang2013communication}. However, the key difference is that the subsamples used by a standard one-shot estimator should not have any overlap with each other. In contrast, the subsamples used by our proposed subsampling methods are allowed to be partially overlapped.

\subsection{Variance and Bias Analysis of the SOS Estimator}
\label{sub:2}
\noindent

To motivate our method, we offer an informal analysis of the bias and variance of the SOS estimator $\thetaSOS$.
The formal theoretical results are provided in Section 2.4. Specifically, by Taylor's expansion, we can approximate $\widehat \theta^{(k)}$ as
$$
\label{eq:ty for thetak}
\widehat{\theta}^{(k)} \approx \theta+\dot{g}(\mu) \left(\widehat \mu^{(k)}-\mu\right)+\frac{1}{2}\ddot{g}(\mu)\left(\widehat \mu^{(k)}-\mu\right)^{2},
$$
where $\dot{g}(\mu)$ and $\ddot{g}(\mu)$ are the first- and second-order derivatives of $g(\mu)$ with respect to $\mu$, respectively. Accordingly, we have
\beq
\label{eq:taylor for bartheta}
\thetaSOS = \frac{1}{K} \sum_{k=1}^K  \widehat{\theta}^{(k)} \approx \theta+\frac{\dot{g}(\mu) }{K} \sum_{k=1}^{K}\left(\widehat \mu^{(k)}-\mu\right)+ \frac{\ddot{g}(\mu)}{2K} \sum_{k=1}^{K}\left(\widehat \mu^{(k)}-\mu\right)^{2}.
\eeq
By equation (\ref{eq:taylor for bartheta}) we known that $\operatorname{var} (\thetaSOS)$ can be approximated by the variance of $\dot{g}(\mu) K^{-1} \sum_{k=1}^{K}\left(\widehat \mu^{(k)}-\mu\right)$. Let $\overline{X} =  N^{-1} \sum_{i\in \mathbb{S}} X_i$ and  $\widehat{\sigma}^2 = N^{-1} \sum_{i\in \mathbb{S}} (X_i - \overline{X})^2 .$ With a slight abuse of notation, we use $\mathbb{S}$ to represent the information contained in the whole sample, that is, the $\sigma$-field generated by $\{X_1,\dots,X_N\}$. Recall that, conditional on $\mathbb{S}$, $X_i$s are independent and identically distributed for any $i \in \mathcal{S}_k$ and $ 1 \leq k \leq K$. We then have $E (\widehat{\mu}^{(k)} | \mathbb{S})= n^{-1} E \left(\sum_{i\in \mathcal{S}_k} X_i | \mathbb{S}\right) = \overline{X}$ and $\operatorname{var} \left(\widehat{\mu}^{(k)} | \mathbb{S}\right)= n^{-2} \operatorname{var} \left(\sum_{i\in \mathcal{S}_k} X_i  | \mathbb{S}\right) = n^{-1}\widehat{\sigma}^2$. Assume that the second moment of $X_1$ is finite, with $\sigma^2 = \operatorname{var} (X_1)$. We then have
\begin{equation}
\label{eq:EV}
\begin{split}
E \left\{\operatorname{var}\left(\thetaSOS| \mathbb{S}\right) \right\} \approx & {} \ \   \frac{\dot{g}(\mu)^2}{K} E \Big\{ \operatorname{var} \left(\widehat{\mu}^{(k)}|\mathbb{S} \right) \Big\} \approx \frac{\dot{g}(\mu)^2}{nK}
\sigma^2, \\
\operatorname{var} \left\{ E\left(\thetaSOS|\mathbb{S}\right) \right \}\approx & {} \ \  \dot{g}(\mu)^2 \operatorname{var} \left( \bar{X} - \mu \right) = \frac{\dot{g}(\mu)^2}{N}  \sigma^2.
\end{split}
\end{equation}
By equation (\ref{eq:EV}), we find that $\operatorname{var} (\thetaSOS)$ can be approximated by $\tau_1\{1/(nK)+1/N\}$, with $\tau_1 = \dot{g}(\mu)^2 \sigma^2.$ Under the condition $nK \gg N,$ we then determine that the variance of the subsample estimator can be further approximated by $\tau_1/N$, which is the asymptotic variance of the WS estimator $\widehat{\theta}$.

Next, we study the bias of $\thetaSOS$. We define the bias of $T_n$ as $\text{Bias} (T_n) = E(T_n) - \theta $ for any estimator $T_n$ of $\theta$. Then, by equation (\ref{eq:taylor for bartheta}), we have
\begin{equation} \label{eq:ty for bias}
\begin{split}
\operatorname{Bias} (\thetaSOS)  &= E \left(\frac{1}{K} \sum_{k = 1}^K \widehat{\theta}^{(k)}\right) -\theta  =  E( \widehat{\theta}^{(k)} ) - \theta  \approx \frac{\ddot{g}(\mu)}{2n}   \sigma^2  + \frac{\ddot{g}(\mu)}{2N} \sigma^2.
\end{split}
\end{equation}
The leading term of $\text{Bias}(\thetaSOS)$ is given by $\tau_2/n$, with $\tau_2 = \ddot{g}(\mu) \sigma^2 /2.$ Unfortunately, it does not improve as $K$ increases.
This indicates that the bias of $\thetaSOS$ is of an order $O(n^{-1})$. This is a smaller order term as compared with $ \wh \theta - \theta = O_p(1/\sqrt{N}),$ as long as
$n \gg \sqrt{N}$.
This condition seems to be quite reasonable for a distributed system \citep{huo2015distributed,jordan2019communication}. In that case, $K$ is the number of distributed computers. As a consequence, $K$ is typically much smaller than $n$, where $n$ is the subsample size allocated to each distributed computer. However, this condition could be problematic for a subsampling method. In this case, $K$ is the total number of subsamples and could be very large. In contrast, for computation convenience, the subsample size $n$ could be much smaller than $\sqrt{N}$. This makes the bias introduced in equation (\ref{eq:ty for bias}) possibly non-negligible. To fix this problem, we are motivated to search for an improved estimator for $\theta$ so that its bias can be greatly reduced.  In this regards, jackknife is a  well known method to reduce the bias of estimators \citep{quenouille1949approximate,efron1981jackknife,cameron2005microeconometrics}. However, the performance of Jackknife in subsampling scenario is not clear. This leads to the novel jackknife estimators presented in the next subsection.

\subsection{The Jackknife Estimators}
\noindent

The objective of this subsection is two-fold. The first goal is to develop a jackknife debiased subsample (JDS) estimator for $\theta.$ The second goal is to propose a jackknife standard error (JSE) estimator for the JDS estimator.

First, we develop the JDS estimator to reduce the estimation bias. To this end, we define a jackknife estimator $\widehat\theta^{(k)}_{-j}$ for the $k$th subsample as follows:
$$
\widehat\theta^{(k)}_{-j} = g \left(\widehat{\mu}^{(k)}_{-j} \right), \text{ where }  \widehat{\mu}^{(k)}_{-j} = \frac{1}{n-1} \sum_{ i \in \mathcal{S}_k}^{i \neq j} X_i.
$$
By similar analysis to that for equation (\ref{eq:taylor for bartheta}), we know that $\text{Bias}\big(\widehat\theta^{(k)}_{-j}\big)$ approximately equals $\tau_2/(n-1)$. Then, $n^{-1} \sum_{j \in \mathcal{S}_k} \text{Bias}\big(\widehat{\theta}^{(k)}_{-j}\big) \approx \tau_2/(n-1)$, and $E\big(n^{-1} \sum_{j \in \mathcal{S}_k} \widehat{\theta}^{(k)}_{-j} - $  $ \widehat{\theta}^{(k)}\big) \approx \tau_2/\{n(n-1)\}$. This inspires an estimator for the bias, which is given by $\widehat{\text{Bias}}^{(k)} = (n-1) n^{-1} \sum_{j \in \mathcal{S}_k} \widehat{\theta}^{(k)}_{-j} - (n-1) \widehat{\theta}^{(k)} $. Accordingly, we can propose a bias-corrected estimator for the $k$th subsample as $\thetaJDS^{(k)} = \widehat{\theta}^{(k)} - \widehat{\text{Bias}}^{(k)}.$
Thereafter, $\thetaJDS^{(k)}$s can be further averaged across different $k$. As a consequence, we obtain the final JDS estimator
$\thetaJDS = K^{-1} \sum^K_{k = 1} \thetaJDS^{(k)}$. Subsequently, we rigorously verify that $\text{Bias}(\thetaJDS)$ is much smaller than that of $\thetaSOS.$ Specifically, $\text{Bias}(\thetaJDS) = O(1/n^2) + O(1/N) $ and $\text{Bias} (\thetaSOS )\approx \tau_2/n$; see equation (\ref{eq:ty for bias}). Furthermore, we can theoretically prove that the asymptotic variance of $\thetaJDS$ remains the same as that of the WS estimator. As a consequence, assuming that $K$ is large enough, excellent statistical efficiency can be achieved by $\thetaJDS$ with a very small subsample size $n$.

Other than bias correction, the jackknife method can also serve as an excellent estimator for the standard error, that is, the standard deviation of the JDS estimator $\thetaJDS$. The basic idea is given as follows. Recall that by equation (\ref{eq:EV}), we know that
\beq
\label{eq:jackSE1}
\operatorname{var}(\thetaSOS) \approx \dot{g}(\mu)^2 \sigma^2 \left(\frac{1}{nK} +\frac{1}{N} \right).
\eeq
Because $N,n$ and $K$ are all known to the user, the key objective here is to estimate the unknown parameter $\tau_1 = \dot{g}(\mu)^2 \sigma^2$. Moreover, by the definition of the jackknife estimator and Taylor's expansion, we have
$$
\widehat{\theta}^{(k)}_{-j} - \widehat{\theta}^{(k)} \approx \dot{g}(\mu) \left(\widehat \mu^{(k)}_{-j}-\widehat \mu^{(k)}\right) =   \frac{\dot{g}(\mu) }{n-1} \left(\widehat \mu^{(k)} - X_j \right)
$$
for any $j \in \mathcal{S}_k$ and $1 \leq k \leq K$. We know immediately that
$$
\label{eq:jackSE2}
E\left\{ \left(\widehat{\theta}^{(k)}_{-j} - \widehat{\theta}^{(k)} \right)^2  \right\} = E \left[ E\left\{ \left(\widehat{\theta}^{(k)}_{-j} - \widehat{\theta}^{(k)} \right)^2  \Big| \mathbb{S} \right\} \right]  \approx \frac{\dot{g}(\mu)^2 \sigma^2}{n(n-1)} = \frac{\tau_1}{n(n-1)},
$$
which is closely related to the unknown parameter $\tau_1$ in equation (\ref{eq:jackSE1}). This is because the sample mean of $\left(\widehat{\theta}^{(k)}_{-j} - \widehat{\theta}^{(k)} \right)^2$
across different $j$ and $k$ is a reasonable approximation of $E\left\{ (\widehat{\theta}^{(k)}_{-j} - \widehat{\theta}^{(k)} )^2  \right\}$. This inspires the following JSE estimator $\widehat{\mathrm{SE}}$:
$$
\widehat{\mathrm{SE}}^2 = \left(\frac{1}{K} + \frac{n}{N} \right) \frac{1}{K} \sum_{k=1}^K  \sum_{j\in \mathcal{S}_k} \left( \widehat{\theta}^{(k)}_{-j} -  \widehat{\theta}^{(k)} \right)^2.
$$
We will theoretically prove that $\widehat{\mathrm{SE}}^2$ is a consistent estimator of $\operatorname{var}(\thetaJDS).$ In addition, $\operatorname{var}(\thetaJDS)/\operatorname{var}(\thetaSOS) = 1+o(1).$ Consequently, $\widehat{\mathrm{SE}}^2$ is also a consistent estimator of $\operatorname{var}(\thetaSOS)$.

\subsection{Theoretical Properties}
\noindent

In this subsection, we study the theoretical properties of the three estimators (i.e., the SOS, JDS and JSE estimators).  To this end, the following standard technical conditions are needed.

\begin{itemize}
\item[(C1)] ({\sc Sub-Gaussian Distribution}) Assume $X_i$ follow a sub-Gaussian distribution, i.e., there exists positive constants $C,\nu$ such that $P(|X_i| > t) \leq C \exp\{-\nu t^2\}$ for every $t > 0$.

\item[(C2)] ({\sc Smoothness condition})      Define $g^{(k)}(\cdot)$ as the $k$th order derivative function of $g(\cdot)$ and assume  $g^{(k)}(\cdot)$ is a continuous function for $k\le 8$.

\item[(C3)] ({\sc Subsampling condition}) As $N \to \infty,$  the subsample size $n \to \infty$. In addition, assume that $n < N, N = o(n^4)$ and $\log K = o(\sqrt{n}).$
\end{itemize}
\noindent
Condition (C1) is a classical and flexible assumption on covariates \citep{jordan2019communication,screening2021zhu}. Condition (C2)  requires the $g$-function to be sufficiently smooth so that a Taylor's expansion can be obtained around $\mu.$  
We require slightly stronger condition since we will derive the asymptotic bias in more explicit forms. The condition can be relaxed to requiring $g(\cdot)$ to be fourth 
continuously differentiable function to guarantee the asymptotic normality
\citep{wu1986jackknife,lehmann2006theory}. 
Lastly, Condition (C3) states the relationships between $n,N$ and $K$.
It requires that the subsample size should 
be large enough to facilitate the asymptotic analysis of higher order terms.
In addition, we require $\log K = o(\sqrt{n})$ to guarantee a uniform convergence
for all subsamples, which is easy to satisfy in practice.

We next consider how to understand the asymptotic behavior of various subsample estimators without a finite moment constraints. Inspired by the asymptotic theory of \cite{shao2003mathmetical}, we adopt here a Taylor's expansion approach. Specifically, take $\thetaSOS$ as an example, by the Taylor's expansion, we have $\thetaSOS = K^{-1} \sum_{k=1}^K \widehat{\theta}^{(k)} = K^{-1} \sum_{k=1}^K  g(\widehat{\mu}^{(k)}) = \theta + \DeltaSOSone + \DeltaSOStwo + \mcO$,
where $\DeltaSOSone =  \dot{g}(\mu) K^{-1} \sum_{k=1}^K (\wh \mu^{(k)} - \mu)$, $ \DeltaSOStwo  = K^{-1} \sum_{k=1}^K \big\{ \ddot{g}(\mu)$  $ (\widehat{\mu}^{(k)} - \mu)^2/2 + \dddot{g}(\mu) (\widehat{\mu}^{(k)} - \mu)^3/6 \big\} ,$ and
$\mcO$ stands for higher order terms. 
As we have discussed informally in Section 2.2, it suggests that the asymptotic behavior of $\thetaSOS - \theta$ could be fully determined by $\DeltaSOSone$ and $\DeltaSOStwo$. Here $\DeltaSOSone$ is unbiased and mainly contributes the variance, while $\DeltaSOStwo$ has ignorable variance and mainly controls the bias.
We can accordingly understand the asymptotic performance of $\thetaSOS$'s bias  by $E(\DeltaSOStwo)$ and $\thetaSOS$'s variance by $\var(\DeltaSOSone)$. Specifically, we have the following theorem.

\begin{theorem}
\label{thm:SOS}
Assume conditions (C1)-(C3) hold, then we have
$
\thetaSOS - \theta = \DeltaSOSone + \DeltaSOStwo + \mcO
$
with $E(\DeltaSOSone) = 0,$ $\var(\DeltaSOStwo) = o\big\{ 1/(nK) + 1/N \big\}$, $\mcO =  o_p\big(1/n + 1/N + \sqrt{1/(nK) + 1/N}\big),$ and
\beqr
\label{eq:thm1.1}
E(\DeltaSOStwo) &=&   \tau_2 \left( \frac{1}{n} + \frac{1}{N} \right) + o\left( \frac{1}{n} \right) \\
\label{eq:thm1.2}
\var(\DeltaSOSone) &=& \tau_1 \left( \frac{1}{nK} + \frac{1}{N} \right) + o\left( \frac{1}{nK} + \frac{1}{N}  \right),
\eeqr
\end{theorem}
\noindent
By Theorem \ref{thm:SOS}, first we  find that the higher order terms $\mcO$ could be  ignorable compared with $\DeltaSOSone$ and $\DeltaSOStwo$. In addition, the asymptotic bias behavior of $\thetaSOS$ is decided by  $\DeltaSOStwo$, while  the asymptotic variance behavior of $\thetaSOS$ is determined by $\DeltaSOSone$. Then by equation (\ref{eq:thm1.1}), we know that the bias of $\DeltaSOStwo$ is affected by both $N$ and $n$. The $1/n$ and $1/N$ terms represent the asymptotic bias due to the subsampling and overall sampling errors, respectively. The leading term of variance for  $\DeltaSOSone$ also includes two quantities. They are the $1/(nK)$ and $1/N$ terms. The first term is due to the subsampling error, and the second term is due to the overall sampling error. Recall that the asymptotic variance of the WS estimator $\widehat{\theta}$ approximately equals $\tau_1/N.$ Then, for the SOS estimator to achieve the same asymptotic efficiency as $\widehat{\theta},$ we must have $nK/N \to \infty.$  Unfortunately, the subsampling error term of $\operatorname{Bias}(\DeltaSOStwo)$ is $O(1/n),$ which does not reduce at all as $K \to \infty.$
Consequently, we need to have $n\gg \sqrt N$ so that the asymptotic bias is  of $o(1/\sqrt{N})$. Otherwise, the SOS estimator can never be asymptotically as efficient as the whole sample estimator $\widehat{\theta}.$
Similar with $\thetaSOS$, we could express $\thetaJDS$ by the Taylor's expansion as $\thetaJDS = \DeltaJDSone + \DeltaJDStwo + \mcO,$ the detailed expression is given in Appendix B. Define $\tau_3  = \dddot{g}(\mu) \mu_3/6, \tau_4 = \ddddot{g}(\mu) \sigma^4 /8,$ and ${\mu}_3 = E (X_i - \mu)^3.$   We next analyze the properties of the JDS estimator in the following theorem.

\begin{theorem}
\label{thm:SJD}
Assume conditions (C1)-(C3) hold, then we have  $
\thetaJDS - \theta = \wh \Delta^{(1)}_{_{JDS}} + \DeltaJDStwo + \mcO
$ with $E(\DeltaJDSone) = 0,$ $\var(\DeltaJDStwo) = o\big\{ 1/(nK) + 1/N \big\}$, $\mcO = o_p\big(1/n^2 + 1/N+\sqrt{1/(nK) + 1/N}\big)$ and
\beqr
\label{eq:thm2.1}
E(\DeltaJDStwo) &=& \frac{\tau_2}{N}  + \frac{\tau_3 + \tau_4}{n^2} + o\left(\frac{1}{N} + \frac{1}{n^2} \right) \\
\label{eq:thm2.2}
\var(\DeltaJDSone) &=& \tau_1 \left( \frac{1}{nK} + \frac{1}{N} \right) + o\left( \frac{1}{nK} + \frac{1}{N} \right) .
\eeqr
\end{theorem}
\noindent
Comparing (\ref{eq:thm1.1}) and (\ref{eq:thm2.1}), we find that for the JDS estimator, the bias term due to the subsampling error is substantially reduced. It is only of the order $1/n^2.$ In contrast, that of the SOS estimator is much larger and is of the order $1/n.$ 
Comparing (\ref{eq:thm1.2}) and (\ref{eq:thm2.2}), we conclude that the leading terms for the variance of both estimators are identical. They can be consistently estimated by the proposed JSE estimator $\widehat{\operatorname{SE}}$. Its asymptotic property is given as follows.

\begin{theorem}
\label{thm:SJSE}
Define $\tau^2 = \tau_1 \left\{1/(nK) + 1/N \right\},$ and further assume conditions (C1)--(C3) hold. The JSE estimator is then ratio consistent for $\tau$, that is, $\widehat{\operatorname{SE}}^2 / \tau^2 \rightarrow_p 1,$ where ``$\rightarrow_p$'' stands for ``convergence in probability''.

\end{theorem}

Lastly, for valid asymptotic inference, we need to study the asymptotic distributions of the JDS estimator $\thetaJDS$ and the SOS estimator $\thetaSOS.$ Consequently, we develop the following theorem to establish  the asymptotic normality for both $\thetaJDS$ and $\thetaSOS.$

\begin{theorem}
\label{thm:normal}
Assume conditions (C1)-(C3) hold.  The JDS estimator $\thetaJDS$ is then asymptotically normal with $
\big( \thetaJDS - \theta \big)/ \tau \rightarrow_d N(0,1),
$
where ``$\rightarrow_d$'' represents ``convergence in distribution''. If one can impose the stronger condition that $n/N^{1/2} \to \infty,$ then the SOS estimator $\thetaSOS$ is also asymptotically normal with $
\big( \thetaSOS - \theta \big)/ \tau \rightarrow_d N(0,1).
$
\end{theorem}
\noindent
From Theorem \ref{thm:normal}, we know that both the SOS and JDS estimators are asymptotically normal. However, the technical conditions required by both estimators are different. The JDS estimator requires $n/N^{1/4} \to \infty.$ This is a condition that can be very easily satisfied. However, for the SOS estimator, a much stronger condition (i.e., $n/N^{1/2} \to \infty$) is required \citep{huo2015distributed,jordan2019communication,wangfei2020efficient}.


\section{NUMERICAL ANALYSIS}
\subsection{Why Sampling with Replacement}
\noindent

We aim to develop a GPU-based algorithm for the proposed method with data being placed on the hard drive. Thus, it is important to understand the sampling mechanism on the hard drive. In particular, we want to carefully elaborate computational efficiency between different sampling mechanisms (i.e. simple random sampling with replacement and simple random sampling without replacement on the hard drive) in the following steps.

\begin{figure}[h]
\centering
\setlength{\abovecaptionskip}{1pt}
\subfigure{
\includegraphics[width=0.9\columnwidth]{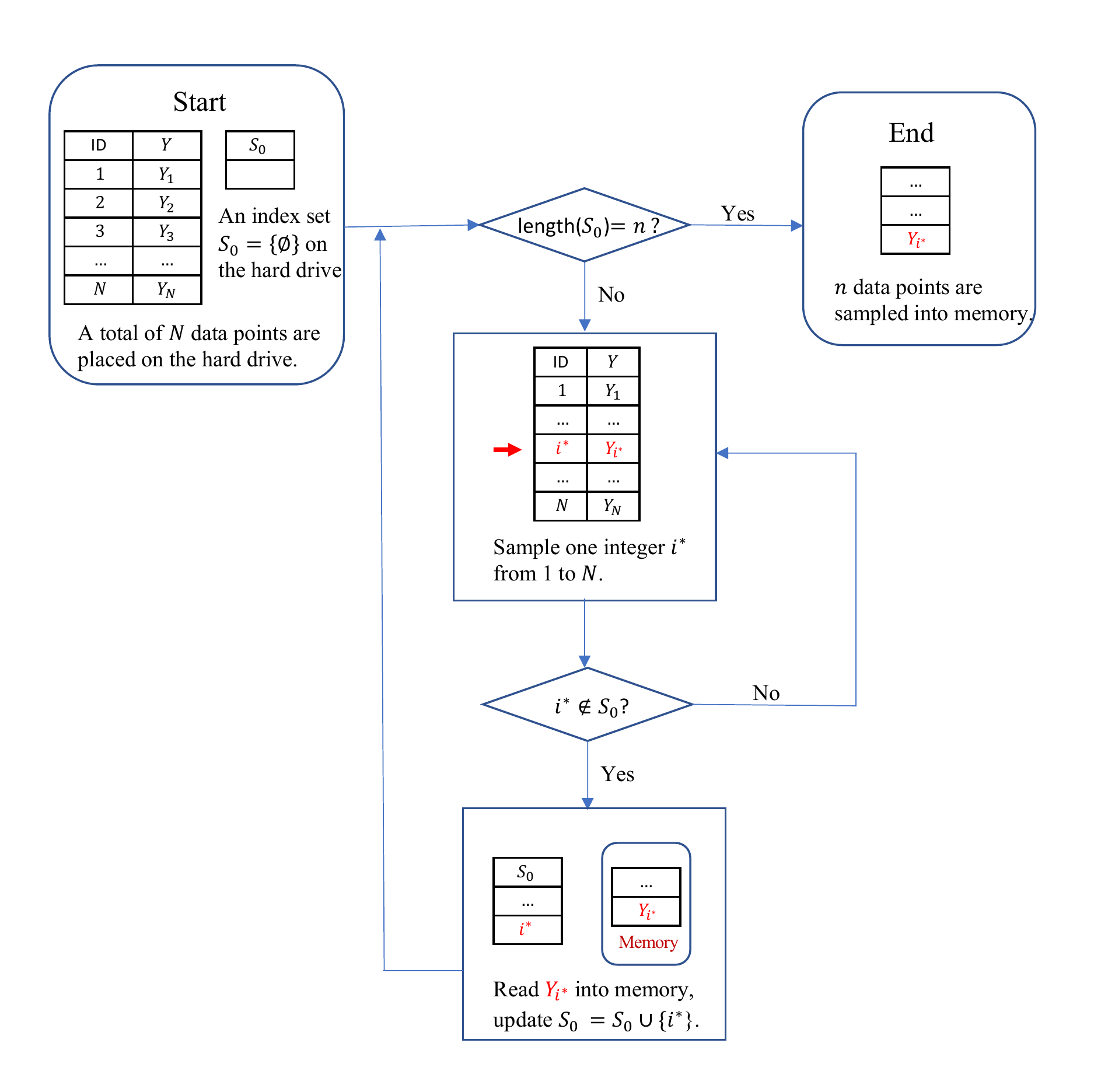} }
\caption{The whole process for sampling without replacement on the hard drive.}
\label{fig:ad1}
\end{figure}

\begin{itemize}
\item[(1)] First, we assume that there are a total of $N$ data points (representing a massive dataset) placed on the hard drive. They are displayed in the top left of Figure \ref{fig:ad1}. It contains a total of $2$ columns. The first column is the sample ID ($n=1,2,3,4,5,\dots,N$) and the second column is the interested variable $Y = (Y_1,\dots, Y_N)$, which represents the interested information.
\item[(2)] Second, to conduct random sampling, we can randomly generate an integer $i^*$ between $1$ and $N$.  This determines which data line should be sampled. Without loss of generality, assume that the sampled unit is $i^*$. Then we read $Y_2$ into memory. (We should note that this sampling procedure is a simplified version. In practice, we cannot access a data line by its sample ID on the hard drive. Instead, we refer to it according to its physical address on the hard drive. This is also not a very straightforward operation and fairly sophisticated). We then update the index set $\mS_0$ from $\mS_0 = \{\emptyset\}$ to $\mS_0 = \{i^*\}.$
\item[(3)] Third, we should explain how to conduct random sampling without replacement. To this end, we randomly and independently generate another integer $i_2^*$ from $1$ to $N$. It is possible that $i_2^*$ is an already sampled unit in $\mS_0$, which leads to duplicated sampling. To avoid duplicated sampling, $i_2^*$ needs to be compared with every already sampled unit in $\mS_0$. If we find $i_2^* \in \mS_0$ already, then $i_2^*$ needs to be re-generated. Otherwise, $\mS_0$ can be updated to be $\mS_0:=\mS_0\cup \{i_2^*\}$ and $Y_{i_2^*}$ is read into memory.

\item[(4)] Assume a total of $K$ subsamples with size $n$ needs to be generated. Then, the size of the index set is about $|\mS_0| =O(nK)$. To avoid duplicated sampling, every sampled unit needs to be compared with every unit in $\mS_0$. This leads to a computation cost of order $O(nK)$ for every sampled unit on average. The total computation cost should be of order $O\big\{(nK)^2\big\}$ on average. This is an expensive cost. The whole process is graphically illustrated in Figure \ref{fig:ad1}.

\item[(5)]
Lastly, if we conduct random sampling with replacement, we \textbf{avoid} the need to: (a) keep updating $\mS_0$ and compare whether $i^*_2 \in \mS_0$; or (b) keep updating $\mS_1$. This makes our proposal computationally more efficient.
\end{itemize}

To summarize, compared with subsampling with replacement, subsampling without replacement with massive datasets is practically
challenging.  Therefore, for massive datasets on a hard drive we prefer sampling
methods with replacement. In this case, no recording and comparison operations need to be conducted. Next, to further demonstrate this point, we develop an experiment to compare sampling with and without replacement on the hard drive. 
To this end, we generate independent and identical $X_i = (X_{i_1}, X_{i_2})$ from a standard bivariate normal distribution with $N = 10^9$. 
The interested parameter is the population mean $\mu$.  To estimate $\mu,$ the sample mean is calculated based on the two sampling strategies. We denote $\wh \mu_{rep}$ and $\wh \mu_{worep} $ to  represent the estimator based on sampling with replacement and without replacement, respectively.
We repeat the experiment $R = 100$ times. Then, the average mean square error (MSE) and time cost (TC) for sampling are reported for both
sampling strategies across $R$ replications.
All the results are summarize in Table \ref{tab:samp}.

\begin{table}[h]
\renewcommand\arraystretch{1.3}
\centering
\caption{\label{tab:samp} Comparison of the sampling with and without replacement on the hard drive based on $R=100$ simulation replications for various $(n,K)$ combinations.}
\begin{spacing}{1.0}
\setlength{\tabcolsep}{6mm}{
\begin{tabular}{cc|cc|cc}
\hline
\hline
\multirow{2}{*}{$n$}&
\multirow{2}{*}{$K$}& \multicolumn{2}{c|}{TC} & \multicolumn{2}{c}{MSE ($ \times 10^{-4}$)}   \\
&&$\wh \mu_{rep}$&$\wh \mu_{worep}$&
$\wh \mu_{rep}$&$\wh \mu_{worep}$ \\
\hline
100&50&0.18&0.49&$4.04$&$4.04$\\
&100&0.36&1.55&$2.03 $&$2.31$\\
&200&0.73  &5.41&$0.87 $ & $0.90$\\

\hline
500&50&0.43&7.64 &$0.81 $ &$0.79 $ \\
&100&0.87 &28.80 &  $0.38 $& $0.36$\\
&200&1.73& 110.97&$0.18 $&$0.19$ \\
\hline
\end{tabular}}
\end{spacing}
\end{table}

From Table \ref{tab:samp}, we draw the following
conclusions. First, the MSE values of the two sampling strategies are comparable, and they both decrease with increasing $n$ or $K$.
However, the TC values of the two strategies are quite different. Sampling with replacement method is much faster than the sampling without replacement method.
As $nK$ increases, the gap between the two strategies increases significantly. For instance, if $n = 500$ and $K = 200$, it takes only $1.73$  seconds for the sampling with replacement method to complete the procedure, while the time required by the sampling without replacement method is almost $111$ seconds.

\subsection{An Algorithm for a GPU}
\noindent

We next develop a GPU-based algorithm for fast computation. Note that the proposed method exhibits many theoretically and practically useful properties. Theoretically, it guarantees the statistical efficiency of subsample estimators with small subsample sizes. Practically, it is simple, automatic, and flexible. However, the associated computation cost is expensive because the new method requires not only subsampling $K$ times but also jackknifing $n$ times for each subsample. Consequently, it is computationally expensive. Accordingly, its implementation on a Central Processing Unit (CPU) might be inefficient because a standard CPU usually has a very limited number of computation cores. To ameliorate this issue, consider for example the MacBook Pro (13-inch, 2020). It uses the Intel Core i5 processor with only four cores. In contrast, a standard Graphical Processing Unit (GPU) may hold tens of thousands of cores.
Accordingly, the GPU is an extremely powerful tool for parallel computation \citep{kruger2005linear,Che2008cuda}. Meanwhile, our method (particularly the jackknifing part) is extremely suitable for parallel computation. This inspires us to develop a GPU-based algorithm for the proposed method.

A standard GPU system should have two unique features. To make full use of its computational power, we need to take both features into consideration.
\begin{figure}[h]
\centering
\includegraphics[width=1\textwidth]{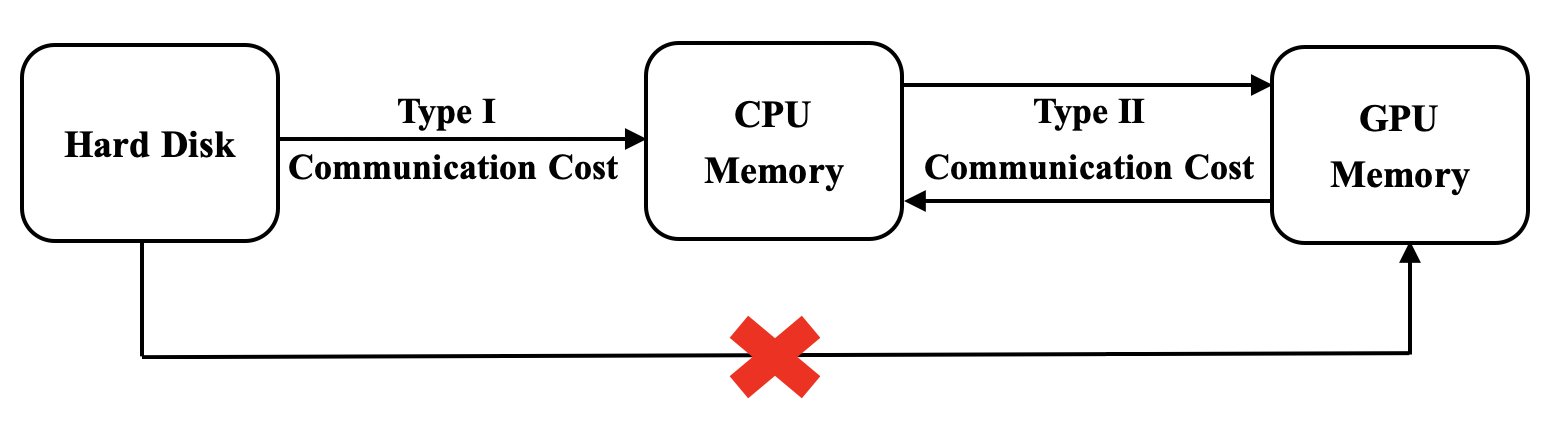}
\caption{Two types of communication cost for a GPU system.}
\label{fig:Sec3.1}
\end{figure}
The first unique feature of the GPU system is that it suffers from two types of communication cost; see Figure \ref{fig:Sec3.1}. The first type of communication cost refers to the time cost required for transferring data from the hard disk (HD) to the CPU memory (CM). This is a standard communication cost that is essentially required by any computation system. For our algorithm, this type of cost is primarily due to subsampling. The second type of communication cost refers to the time cost required for transferring data from the CM to the GPU memory (GM). The main purpose of transferring data from the CM to GM is to prepare data for parallel execution of jackknifing. Consequently, we consider that this part of the communication cost is mainly due to jackknifing. Note that the current GPU architecture does not allow the GPU to directly read the data from the HD. As a consequence, a good algorithm should simultaneously minimize both types of communication cost. Multiple communication between the HD, CM and GM should be avoided.

\begin{figure}[h]
\centering
\includegraphics[width=1\textwidth]{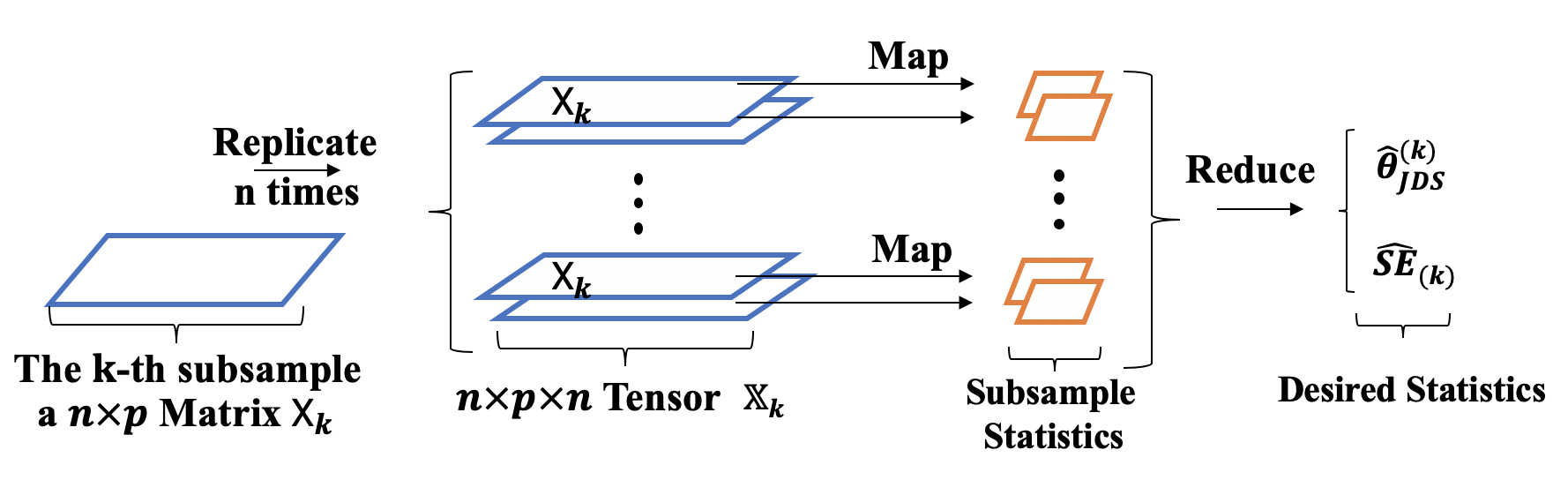}
\caption{A graphical illustration of the proposed GPU algorithm.}
\label{fig:Sec3.2}
\end{figure}

The second unique feature is that GPU systems are extremely suitable for tensor-type parallel computation. Through this type of computation, the parallel computation power of a GPU system can be fully utilized. This suggests that the jackknifing computation should be formulated into a tensor-type computation problem. Specifically, here, we develop a three-step algorithm to implement the proposed method. The process is shown in Figure \ref{fig:Sec3.2}. First, we obtain the $k$th subsample $\mathcal{S}_k \in \mathbb{S}$ from the HD and place it in the CM. With slight abuse of notation, we assume that in this subsection, $X_i$ is a $p$-dimensional vector for any $i \in \mathbb{S}$. Next, we can formulate the $k$th subsample into an $n \times p$ matrix format as $\mathsf{X}_k = [X_i,i\in S_k] \in \mathbb{R}^{n \times p}$. We then pass $\mathsf{X}_k$ to the GM and replicate $\mathsf{X}_k$ $n$ times so that a 3-dimensional tensor $\mathbb{X}_k  = \big[\mathsf{X}_k,\dots,\mathsf{X}_k\big]  \in \mathbb{R}^{n \times p \times n} $ can be constructed.
Next, we define a function to compute the intended statistics with jackknifing. We then map this function to different channels of $\mathbb{X}_k$, where each $\mathsf{X}_k$ represents one channel of $\mathbb{X}_k$. By doing so, jackknifing computation can be executed by the GPU systems in a parallel fashion.
We then collect the computation results from each channel and reduce them into the desired statistics $\widehat{\theta}^{(k)}_{JDS}$ and $\widehat{\operatorname{SE}}^2_{(k)} = \sum_{j \in \mathcal{S}_k} \big( \widehat{\theta}^{(k)}_{-j} - \widehat{\theta}^{(k)} \big)^2$ for the $k$th subsample. We then obtain the final estimators accordingly. This leads to the entire GPU algorithm. The details are provided below.

\begin{algorithm}
\caption{The GPU algorithm}  \label{alg:algorithm1}
\KwIn{ Data $X_1,\dots,X_N$ on the HD, $X_i \in \mathbb{R}^{p}$ \;
\quad\quad\quad\quad $g(\cdot)$: the function of interest of the moment\; \quad\quad\quad\quad $n$: the subsample size \quad\quad\quad\quad $K$: the number of subsamples\;
}
\KwOut{ A JDS estimator $\thetaJDS$ and a JSE estimator $\widehat{\mathrm{SE}}.$ }

\textbf{for} $\ k\leftarrow 1$ $\mathbf{to}$ $K\ $ \textbf{do}  $ \  \text{subsampling}$

\quad  Generate $\mathcal{S}_k \subset \mathbb{S}$, and then, place the $n \times p$ matrix $\mathsf{X}_k$ into the GM \;
\quad  Compute $\widehat{\theta}^{(k)}\leftarrow g(\widehat{\mu}^{(k)})$ in the GM\;
\quad  Generate an $n\times p \times n$ tensor $\mathbb{X}_k$ in the GM \;
\quad  Map the function $g(\cdot)$ to each channel of $\mathsf{X}_k$, which then leads to $\text{ } \{\widehat{\theta}^{(k)}_{-j}, j \in \mathcal{S}_k\}$\;

\quad  Compute $\thetaJDS^{(k)} = \widehat{\theta}^{(k)} -(n-1)\big\{ n^{-1} \sum^{n}_{j=1}\widehat{\theta}^{(k)}_{-j} - \widehat{\theta}^{(k)}\big\}$ and

\quad \quad\quad \quad \quad $\text{ }\widehat{\mathrm{SE}}^{2}_{(k)}=\sum_{j\in \mathcal{S}_k}$ $\big( \widehat{\theta}^{(k)}_{-j} -  \widehat{\theta}^{(k)} \big)^2$ in the GM \;

\textbf{end}

Compute $\thetaJDS = K^{-1} \sum^K_{k = 1} \thetaJDS^{(k)}$ ; $ \widehat{\mathrm{SE}}^2  = (1/K + n/K ) K^{-1} \sum_{k = 1}^K \widehat{\mathrm{SE}}^{2}_{(k)}$ in the GM\;
\Return{$\thetaJDS$ $\operatorname{and}$ $ \widehat{\mathrm{SE}}^2$ }.
\end{algorithm}

\subsection{The Communication and Computation Cost}
\noindent

To evaluate the finite sample performance of the proposed method, we subsequently present a number of numerical experiments. We first consider how to generate the whole sample with a very large $N = 10^{9}.$ For every $1 \leq i \leq N$, we generate a 2-dimensional random variable $X_i = (X_{i1},X_{i2})^{\top}$ independently and identically from a bivariate normal distribution with mean 0 and covariance $\Sigma = \{\sigma_{ij}\}_{2 \times 2},$ where $\sigma_{11} = 25, \sigma_{12} = \sigma_{21} = 10$, and $\sigma_{22} = 5.$ We then define the parameter of interest to be the correlation coefficient $\operatorname{Corr}(X_{i1},X_{i2})$ as follows:
$$
\theta = \operatorname{Corr}(X_{i1},X_{i2}) = \frac{\operatorname{Cov} (X_{i1},X_{i2}) }{\sqrt{ \operatorname{var} (X_{i1}) \operatorname{var} (X_{i2}) } } = \frac{2}{\sqrt{5}}.
$$
This parameter is a complex nonlinear function of various moments about $X_i.$ Once the whole sample is generated, it is placed as a single file on the HD, requiring approximately $38.3$ gigabytes. As one can see, this is a size that can hardly be read into a CM. Once the data are placed in the HD, they are fixed for the rest of the simulation experiments. In other words, we do not update the whole sample dataset on the HD across different simulation replications. For a reliable evaluation, we replicate the subsequent experiment a total of $M = 1000$ times. All computations are performed by using TensorFlow 2.2.0 on a single GPU device (NVIDIA Tesla P100).

In this subsection, we focus on the performance in terms of the time cost. We study both the communication cost and computation cost. The communication cost can be further divided into two parts. The first part is the time cost required for transferring data from the HD to the CM. The second part is the time cost required for transferring data from the CM to GM.
Next, we vary the subsample size $n$ from 100 to 3,000 and $K$ from $10$ to $200$. We then use $\mathcal{S}_k^{(m)} \in \mathbb{S}$ to represent the $k$th subsample obtained in the $m$th simulation replication. The time cost used for obtaining $\mathcal{S}_k^{(m)}$ is recorded by $T_{1k}^{(m)}.$
Based on $\mathcal{S}_k^{(m)}$, we can obtain matrix $\mathsf{X}_k^{(m)}.$ We then transfer $\mathsf{X}_k^{(m)}$ from the HD to the CM, where the associated time cost is recorded as $T_{2k}^{(m)}.$
The computation cost required for computing $\thetaJDS$ and $\operatorname{\widehat{SE}}$ is given by $T_{3k}^{(m)}.$ Consequently, the total time cost is given by $T^{(m)}_k =  T_{1k}^{(m)}+ T_{2k}^{(m)} + T_{3k}^{(m)}$. Their averages are obtained as $T_1 = M^{-1} \sum_{k,m} T_{1k}^{(m)},$ $T_2 = M^{-1} \sum_{k,m} T_{2k}^{(m)},$ and $T_3 = M^{-1} \sum_{k,m} T_{3k}^{(m)}.$ Their relationships with both $K$ and $n$ are investigated.

\begin{figure}[h]
\centering
\setlength{\abovecaptionskip}{1pt}
\subfigure{
\includegraphics[width=1\columnwidth]{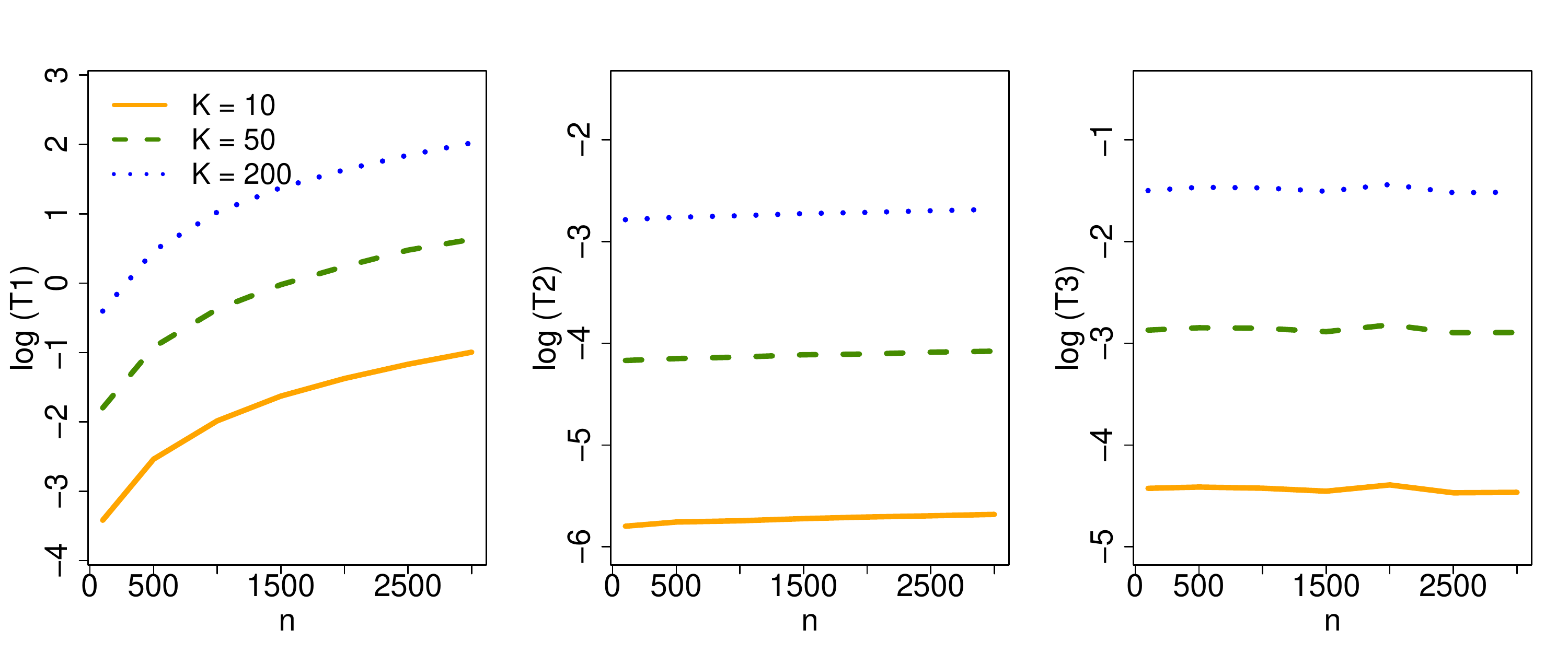} }
\caption{The log-transformed time cost for different $(n,K)$ combinations with $K = 10,50$ and $200$. The communication cost due to subsampling $T_1$ is given in the left panel. The communication cost due to jackknifing $T_2$ is reported in the middle panel. The computation cost $T_3$ is presented in the right panel. The reported time costs (in log-scale) are averaged based on $M = 1000$ simulations.}
\label{fig:1}
\end{figure}
The detailed results are given in Figure \ref{fig:1}. As one can see from Figure \ref{fig:1}, all types of time cost increase as the number of subsamples $K$ increases. In particular, the communication cost required by subsampling (i.e., $T_1$) is substantially larger than the other two types of time cost. Comparatively, the communication cost required by jackknifing (i.e., $T_2$) is the smallest. It is remarkable that  $T_3$ is supposed to be very significant if a CPU-only system is used. However, due to the use of a GPU system, the corresponding time cost becomes practically ignorable. To understand this idea, considering one special case with $K = 50$ and $n = 3000,$ we have $T_1 = 1.882~s, T_2 = 0.017~s$ and $T_3 =0.055~s.$ We also find in the middle and right panels of Figure \ref{fig:1} that for a fixed total subsample size $K$, $T_2$ and $T_3$ remain almost unchanged as $n$ increases. This result demonstrates the excellent parallel capability of a GPU-based system and suggests that better computation efficiency can be achieved by setting the subsample size $n$ to be as large as possible as long as the computer memory allows this.

\begin{figure}[h]
\centering
\setlength{\abovecaptionskip}{1pt}
\subfigure{
\includegraphics[width=0.8\columnwidth]{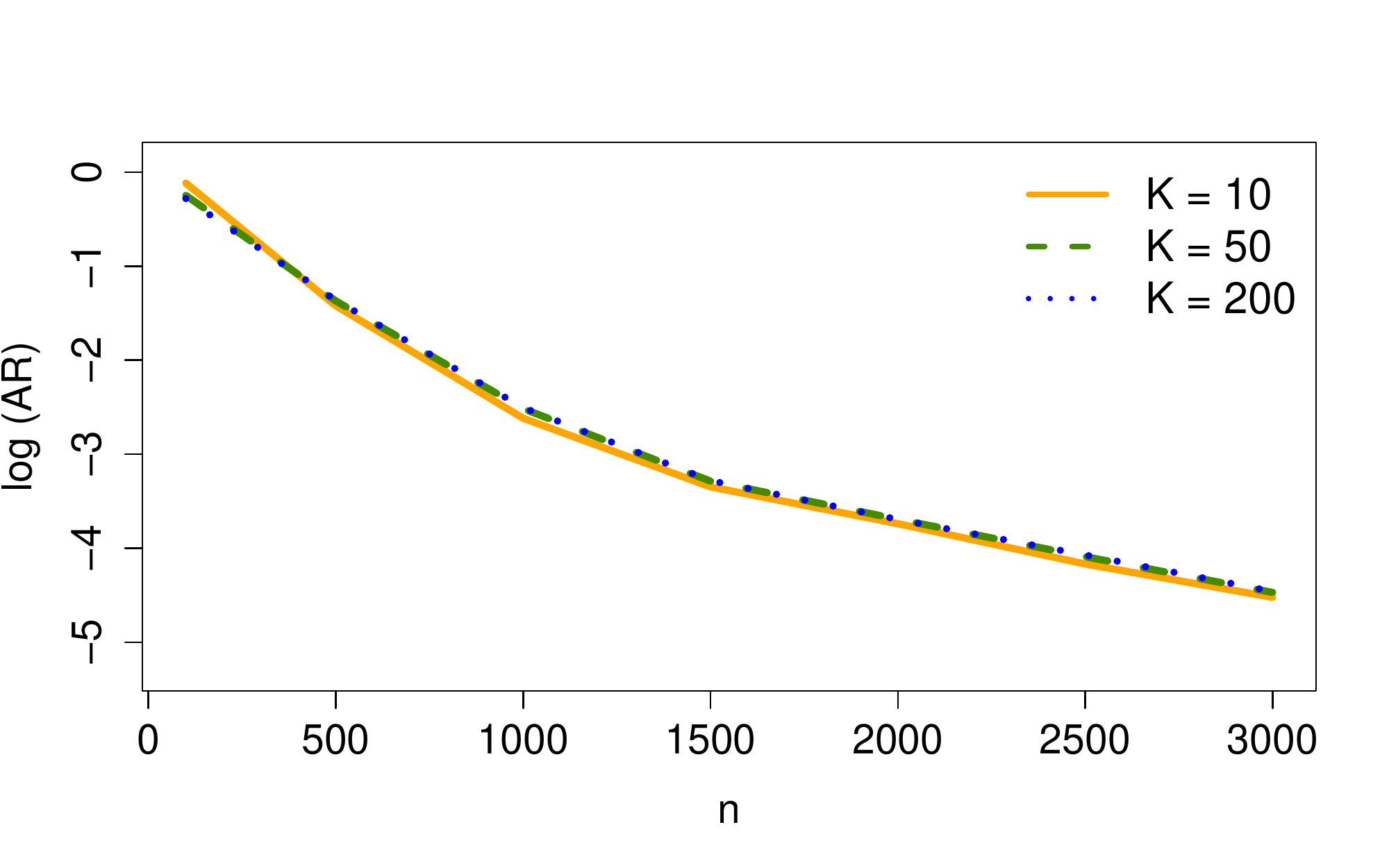} }
\caption{ Comparison of the computation efficiency between a GPU system and a CPU system. The AR is reported in log-scale based on $M=1000$ simulation replications. The numbers of subsamples are fixed to $K = 10, 50$ and $200.$}
\label{fig:2}
\end{figure}

Next, we demonstrate the computational advantage of a GPU system. To this end, we define $T_{_{GPU}}^{(m)}$ as the total time cost required by the $m$th simulation replication except the  communication cost due to subsampling (such cost is required by any computation system). We then execute the same algorithm on a CPU-only system (in our case, TensorFlow 2.2.0 can also be executed on the CPU-only system).  This leads to the total time cost except the  communication cost due to subsampling required by the CPU-only system, which is recorded as $T_{_{CPU}}^{(m)}.$
We then compute their ratio for the $m$th replication as $R^{(m)} = T_{_{GPU}}^{(m)}/ T_{_{CPU}}^{(m)}$. We define the averaged ratio as $\operatorname{AR} = M^{-1} \sum_{m=1}^M R^{(m)}.$ Then, the relationships of the log-transformed $\operatorname{AR}$ values for different $(n,K)$ combinations are reported in Figure \ref{fig:2}. As we can see from Figure \ref{fig:2}, the log(AR) values are always smaller than 0. This suggests that the computational time cost required by a GPU-based system is always smaller than that required by a CPU-only system on average. In fact, the reported log(AR) values seem to be rather insensitive to the number of subsamples (i.e., $K$). Furthermore, for a fixed number of subsamples $K,$ the log(AR) value decreases as the subsample size $n$ increases. This is because a larger $n$ requires a higher computation cost. Accordingly, the parallel computational power of a GPU system can be better demonstrated. For instance, considering the case with $K = 50$ and $n = 3000,$ the averaged time cost of the GPU system is approximately $0.55$ s, while that of the CPU system is approximately $4.83$ s. The corresponding AR value is $\operatorname{AR} = 0.011.$ This suggests that the computational time cost required by a GPU system is only approximately 1.1\% that of a CPU system on average.

\subsection{Simulation Results of the JSE Estimator}
\noindent

\begin{figure}[h]
\centering
\setlength{\abovecaptionskip}{1pt}
\subfigure[$n = 100$]{
\includegraphics[width=0.45\columnwidth]{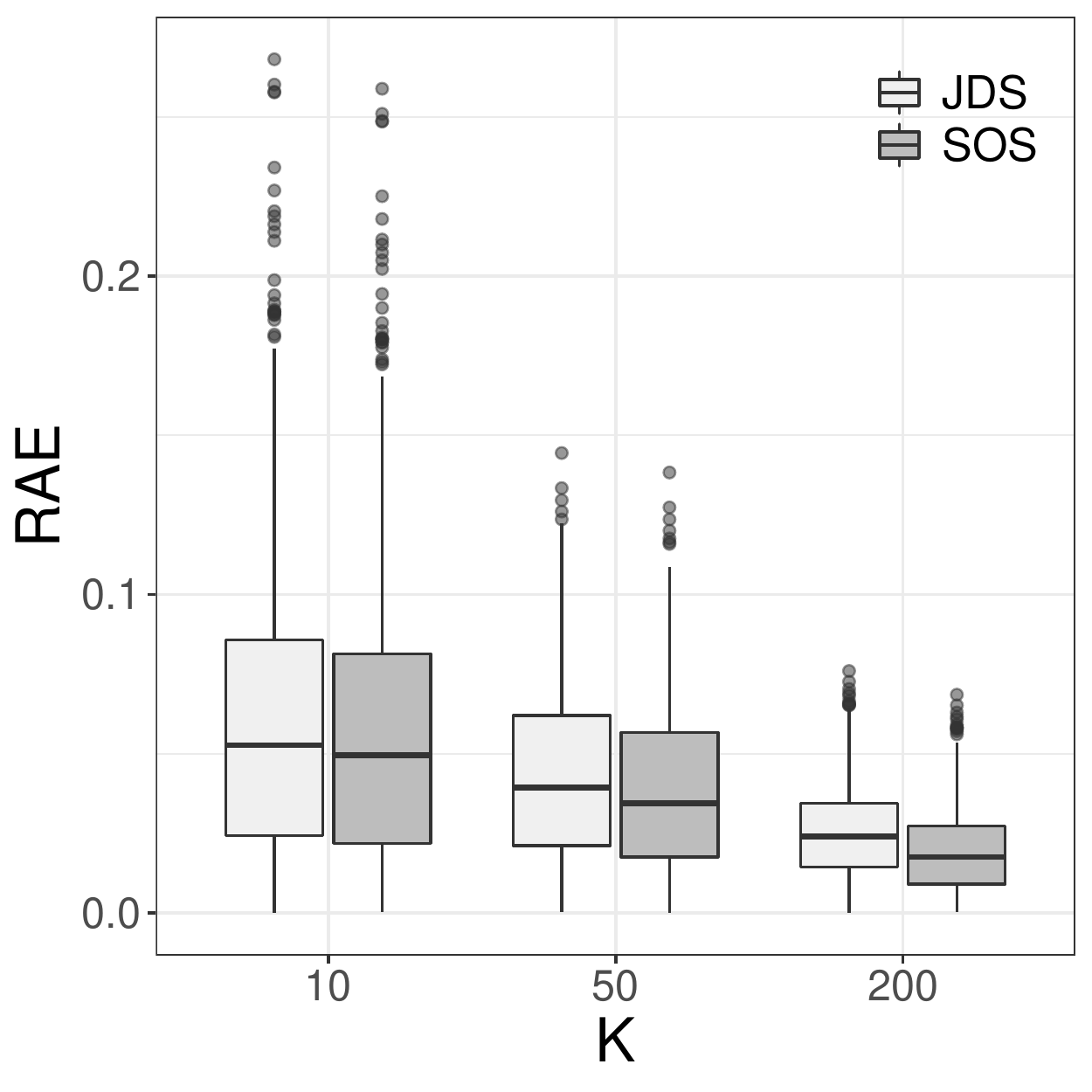} }
\subfigure[$K = 100$]{
\includegraphics[width=0.45\columnwidth]{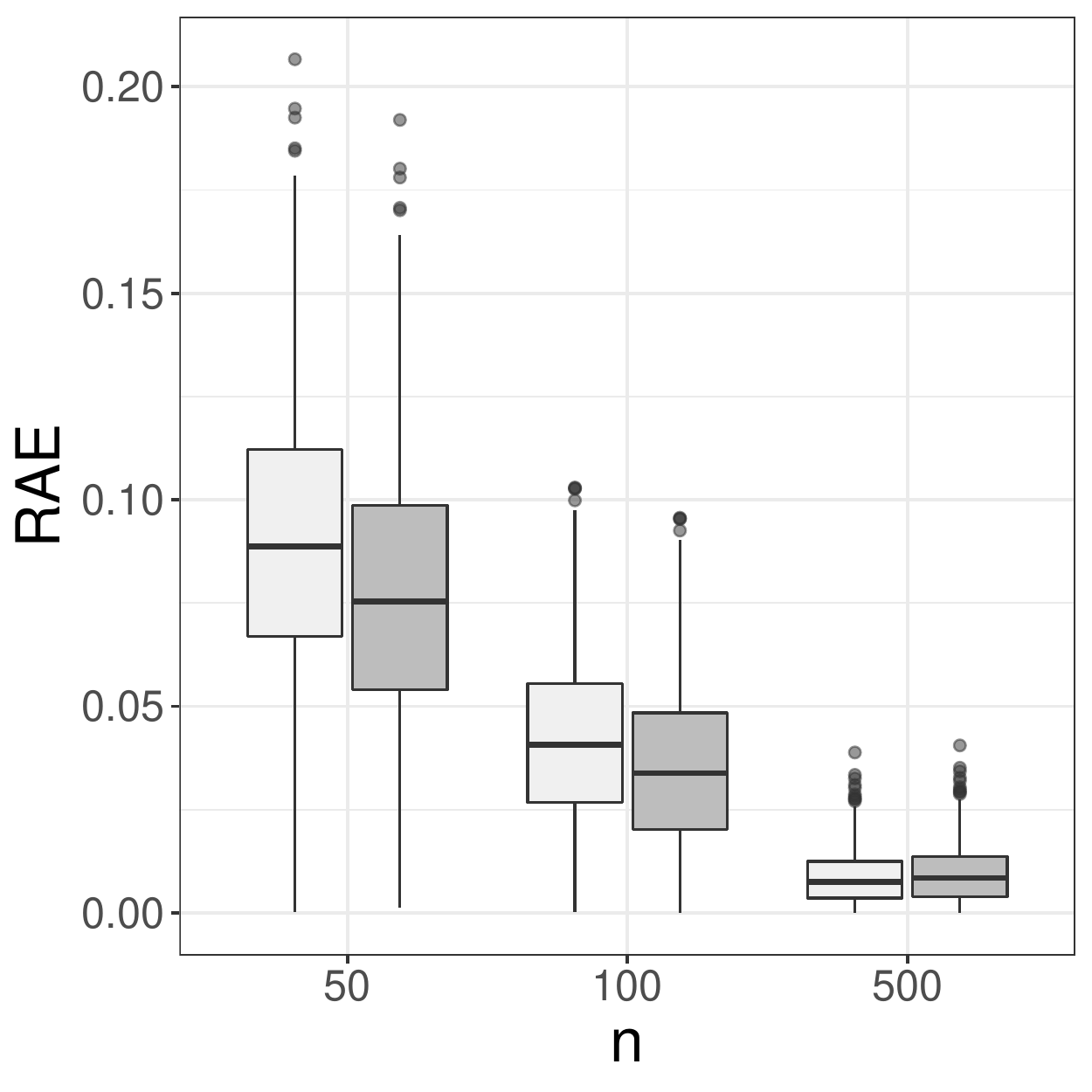} }
\caption{Boxplots of RAE values for the JDS (light box) and SOS (dark box) estimators. The left panel corresponds to the case with $n$ fixed to $n=100$. The right panel corresponds to the case with $K$ fixed to $K =100$. Each box is summarized based on $M=1000$ simulation replications.}
\label{fig:3}
\end{figure}

In this subsection, we focus on the finite sample performance of the JSE estimator $\widehat{\operatorname{SE}}$. To this end, we follow the simulation setup in the previous subsection. Note that the data on the HD are generated only one time to conserve experimental time. Once the data are generated, we replicate experiments $M=1000$ times based on the same whole sample dataset. Specifically, for the $m$th replication, we obtain an SOS estimator $\thetaSOS^{(m)}$, a JDS estimator $\thetaJDS^{(m)},$ and a JSE estimator $\widehat{\operatorname{SE}}^{(m)}$. Define $\operatorname{SE}_{_{SOS}}$ and $\operatorname{SE}_{_{JDS}}$ as the respective sample standard deviations of $\{\thetaSOS^{(m)}, m = 1,\dots,M \}$ and  $\{\thetaJDS^{(m)}, m = 1,\dots,M \}$. Accordingly, $\operatorname{SE}_{_{SOS}}$ and $\operatorname{SE}_{_{JDS}}$ measure the variabilities of $\thetaSOS$ and $\thetaJDS$ conditional on the whole sample dataset on the HD.
Because we have $N \gg nK,$ they should be good approximations of the true variabilities of $\thetaSOS$ and $\thetaJDS$; see Theorems \ref{thm:SJD} and \ref{thm:SJSE}. Next, for the $m$th replication, we can define the relative absolute errors as $\operatorname{RAE}_{_{SOS}}^{(m)} = \big| \widehat{\operatorname{SE}}^{(m)}/ \operatorname{SE}_{_{SOS}} -1 \big| $ and  $\operatorname{RAE}_{_{JDS}}^{(m)} = \big| \widehat{\operatorname{SE}}^{(m)}/ \operatorname{SE}_{_{JDS}} -1 \big|.$
They are box plotted in Figure \ref{fig:3}.

As one can see from the left panel of Figure \ref{fig:3}, the $\operatorname{RAE}$ values of the SOS and JDS estimators are similar. They both decrease to 0 as $K$ increases. This suggests that a larger $K$ leads to more accurate JSE estimators, under the condition that $n$ is fixed. Qualitatively similar patterns are also observed for the right panel. We find that a larger $n$ leads to a more accurate JSE estimation, under the condition that $K$ is fixed. To summarize, both boxplots in Figure \ref{fig:3} suggest that the proposed JSE estimator is consistent as $nK \to \infty$.

\subsection{Simulation Results of the JDS Estimator}
\noindent

Finally, we evaluate the finite sample performance of the point estimation $\thetaJDS$ and its statistical inference in terms of the confidence interval. For comparison, that of the SOS estimator $\thetaSOS$ is also evaluated. Specifically, following the simulation setup in the previous subsection, we replicate experiments $M = 1000$ times based on the same whole dataset on the HD. For the $m$th replication, we calculate the JDS estimator $\thetaJDS^{(m)}$ and the corresponding JSE estimator $\operatorname{\widehat{SE}}^{(m)}.$  This leads to a total of $M$ estimators $\{(\thetaJDS^{(m)}, \widehat{\operatorname{SE}}^{(m)}): 1\leq m\leq M\}.$ Based on these estimators, the averaged bias can be computed as $\operatorname{Bias} = M^{-1} \sum_{m=1}^M \big(\thetaJDS^{(m)} - \theta\big) $, and the corresponding standard error (SE) can be obtained. In addition, for each estimator $\thetaJDS^{(m)},$ a $(1-\alpha)$th level confidence interval for $\theta$ is constructed as $\operatorname{CI}^{(m)} = \big[\thetaJDS^{(m)} - \widehat{\operatorname{SE}}^{(m)}Z_{1-\alpha/2}, \thetaJDS^{(m)} + \widehat{\operatorname{SE}}^{(m)}Z_{1-\alpha/2} \big]$, where $\alpha = 0.05$ and $Z_{\alpha}$ represents the lower $\alpha$ quantile of the standard normal distribution. The empirical coverage probabilities are then also evaluated as $\operatorname{ECP}_{_{JDS}} = M^{-1}\sum_{m=1}^M I(  \theta \in \operatorname{CI}^{(m)} ),$ where $I(\cdot)$ is the indicator function. The SOS estimator $\thetaSOS$ is evaluated similarly. The detailed results are given in Table \ref{tab:model0}.

From Table \ref{tab:model0}, we find that the two estimators perform similarly in terms of the standard error (SE) for various $(n,K)$ combinations. However, they are very different in terms of bias. The bias of the SOS estimator $\thetaSOS$ is much larger than that of $\thetaJDS.$ Considering for example the case with $n = 200$ and $K = 200,$ the bias of $\thetaSOS$ is $4.49\times 10^{-4}$, while that of $\thetaJDS$ is only $1.1\times 10^{-5}.$ As one can see, the former is approximately forty times larger than the latter. Moreover, the bias of $\thetaSOS$ is quite comparable to its standard error. As a consequence, the confidence interval of $\thetaSOS$ is poor, resulting from the fact that the corresponding ECP is significantly smaller than $95\%.$ In contrast, the confidence interval of the JDS estimator is good since the corresponding ECP values of $\thetaJDS$ are quite close to $95\%$.

\begin{table}[h]
\renewcommand\arraystretch{1.3}
\centering
\caption{\label{tab:model0} Comparison of the SOS estimator $\thetaSOS$ and the JDS estimator $\thetaJDS$ based on $M=1000$ simulation replications for various $(n,K)$ combinations.}
\begin{spacing}{0.8}
\setlength{\tabcolsep}{6mm}{
\begin{tabular}{c|cc|cc|cc}
	\hline
	\hline
	\multirow{2}{*}{$K$}& \multicolumn{2}{c|}{SE $(\times 10 ^{-3})$} & \multicolumn{2}{c|}{Bias $(\times 10 ^{-3})$} &  \multicolumn{2}{c}{ECP (\%)}  \\
	&$\thetaSOS$&$\thetaJDS$&
	$\thetaSOS$&$\thetaJDS$&
	$\thetaSOS$&$\thetaJDS$ \\
	\hline
	\multicolumn{7}{c}{$n = 50$} \\
	100&2.964&2.929&2.038&0.050&92.1&95.8\\
	200&2.103&2.077&2.054&0.066&87.8&97.0\\
	500&1.335&1.314&1.956&0.032&73.6&96.8\\
	1000&0.972&0.959&1.907&0.078&50.8&95.4\\
	\multicolumn{7}{c}{$n = 100$} \\
	100&2.068&2.057&0.910&0.029&93.3&96.1\\
	200&1.446&1.437&0.960&0.019&91.7&95.3\\
	500&0.938&0.932&0.877&0.065&84.8&95.6\\
	1000&0.672&0.667&0.904&0.038&72.9&95.6\\
	\multicolumn{7}{c}{$n = 200$} \\
	100&1.436&1.432&0.487&0.029&93.7&94.8\\
	200&1.029&1.025&0.449&0.011&92.9&95.4\\
	500&0.656&0.654&0.442&0.017&89.4&95.0\\
	1000&0.441&0.440&0.477&0.018&83.3&96.4\\
	\hline
	\hline
\end{tabular}}
\end{spacing}
\end{table}

\subsection{Real Data Analysis}
\noindent

In this subsection, we study a real dataset: the U.S. Airline Dataset. The dataset is available on the official website of the American Statistical Association (ASA). The airline dataset contains approximately 120 million records. It takes up approximately 12 gigabytes of space on a hard drive. Each record contains detailed information for one particular commercial flight in the USA from October 1987 to April 2008. The dataset contains 13 continuous variables and 16 categorical variables. For illustration, we focus on the 13 continuous variables. However, a significant portion of records are missing for many continuous variables. Only 5 of them have missing rates less than 10\%: $\mathsf{ActualElapsedTime}$ (actual elapsed time), $\mathsf{CRSElapsedTime}$ (scheduled elapsed time), $\mathsf{Distance}$, $\mathsf{DepDelay}$ (departure delay), and $\mathsf{ArrDelay}$ (arrival delay). As a consequence, only these 5 variables are subsequently illustrated. For more detailed variable information, refer to the ASA official website at \url{http://stat-computing.org/dataexpo/2009}. 

\begin{table}
\centering
\caption{\label{tab:2} Descriptive statistics for the 5 continuous variables based on the whole airline dataset after signed-log-transformation. The descriptive statistics are given by the sample mean (Mean), sample standard deviation (SD) and sample kurtosis (Kurt).}
\begin{spacing}{1}
\begin{tabular}{c|ccccc}
\hline
&$\mathsf{ActualElapsedTime}$&$\mathsf{CRSElapsedTime}$& $\mathsf{Distance}$& $\mathsf{DepDelay}$& $\mathsf{ArrDelay}$\\
\hline
Mean&4.656&4.670&6.272&0.492&0.236\\
SD&0.525&0.513&0.777&1.905&2.463\\
Kurt&2.586&2.597&2.750&2.272&1.594\\
\hline
\end{tabular}
\end{spacing}
\end{table}

For each variable, the signed-log-transformation is applied: $\operatorname{log}|x| \cdot \operatorname{sign}(x)$ transformation. This transformation is conducted purely for illustration. Otherwise, many variables (e.g., $\mathsf{ArrDelay}$) are so heavy-tailed that the existence of finite moments becomes questionable. For each transformed variable, the following parameters are studied: the mean, standard deviation, and kurtosis. Their WS estimators are given in Table \ref{tab:2}. These WS estimators are then treated as if they were the true parameters. Accordingly, simulation experiments can be conducted as in the previous subsections. In this case, we fixed $nK = 6 \times 10^4$, with different $(n,K)$ combinations, and replicated the experiments $M=1000$ times. The detailed results are summarized in Table \ref{tab:3}.

From Table \ref{tab:3}, we can obtain the following interesting observations. First, note that the sample mean is an exactly unbiased estimator for the mean. Accordingly, both the SOS and JDS estimators are unbiased. In fact, they are identical to each other in this case. As a result, both estimators demonstrated identical simulation results, with ECP values both very close to their nominal level of 95\% for all five variables. Second, for the other two parameters (i.e., standard deviation and kurtosis), the sample estimators are no longer unbiased. Accordingly, the SOS and JDS estimators are no longer identical. As we expect, both estimators are similar in terms of the standard error (SE). However, they are very different in terms of the empirical bias. Obviously, the bias of the SOS estimator is substantially larger than that of the JDS estimator for all reported cases. As a consequence, the ECP values of the SOS estimator significantly depart from their nominal level of 95\%. In contrast, those of the JDS estimator remain very close to 95\%. Consider for example the case of the kurtosis of $\mathsf{ArrDelay}$ with $n = 200$ and $K = 300.$ The ECP value of the SOS estimator is only 35.6\%. In contrast, that of the JDS is 95.7\%.

\section{CONCLUDING REMARKS}
\noindent

In this article, we develop a novel statistical method for datasets with large sizes. The new method is particularly designed for practitioners with limited computational resources. The proposed method combines the ideas of both subsampling and jackknifing. Subsampling allows our method to work with datasets with large sizes. Jackknifing further enhances this capability by significantly reducing the bias. To practically implement our method, a novel algorithm is developed for GPU systems. We theoretically show that the resulting estimator could be as good as the whole sample estimator under very mild regularity conditions. Extensive numerical studies built on both simulation and real datasets are presented to demonstrate its outstanding performance.

To conclude this work, we would like to discuss a few interesting topics for future study. First, the statistics considered in this work are relatively simple. They represent nonlinear transformation of various moments. It is then of great interest to develop similar methods for more general $M$ estimators. Second, the data considered in this work are collected from independent samples. This makes the theoretical understanding of the resulting subsample estimator analytically simple. How to develop similar methods for data with a sophisticated dependence structure (e.g., spatial temporal data) is another interesting topic worth studying. Future research along this direction is definitely needed.


\bibhang=1.7pc
\bibsep=2pt
\fontsize{9}{14pt plus.8pt minus .6pt}\selectfont
\renewcommand\bibname{\large \bf References}
\expandafter\ifx\csname
natexlab\endcsname\relax\def\natexlab#1{#1}\fi
\expandafter\ifx\csname url\endcsname\relax
\def\url#1{\texttt{#1}}\fi
\expandafter\ifx\csname urlprefix\endcsname\relax\def\urlprefix{URL}\fi

\bibliographystyle{asa}
\bibliography{ref}

\begin{landscape}
\begin{table}
\centering
\caption{\label{tab:3} Simulation results for the airline dataset based on $M = 1000$ simulation replications. The parameters of interest include the mean (Mean), standard deviation (SD), and kurtosis (Kurt) for signed-log-transformed variables. Both the SOS estimator $\thetaSOS$ and the JDS estimator $\thetaJDS$ are compared in terms of Bias, SE and ECP. The nominal level of ECP is 95\%.}
\begin{spacing}{0.64}
\begin{tabular}{cc|cc|cc|cc|cc|cc}
	\hline
	& & \multicolumn{2}{c}{$\mathsf{ActualElapsedTime}$} & \multicolumn{2}{c}{$\mathsf{CRSElapsedTime}$} & \multicolumn{2}{c}{$\mathsf{Distance}$} & \multicolumn{2}{c}{$\mathsf{DepDelay}$} & \multicolumn{2}{c}{$\mathsf{ArrDelay }$}\\
	\cmidrule(lr){3-4} \cmidrule(lr){5-6} \cmidrule(lr){7-8} \cmidrule(lr){9-10} \cmidrule(lr){11-12}
	& &SOS &JDS &SOS &JDS&SOS &JDS&SOS &JDS&SOS &JDS \\
	\multicolumn{12}{c}{$n = 300$, \quad $K = 200$} \\
	Mean & Bias $(\times 10^{-2})$ & 0.003&0.003&0.002&0.002&0.001&0.001&0.001&0.001&0.026&0.026\\
	& SE $(\times 10^{-2})$ & 0.219&0.219&0.216&0.216&0.326&0.326&0.771&0.771&0.980&0.980  \\
	& ECP (\%)& 94.6&94.6&94.1&94.1&94.1&94.1&94.6&94.6&95.7&95.7 \\
	
	\multicolumn{12}{c}{} \\
	
	SD & Bias $(\times 10^{-2})$& 0.130&0.008&0.127&0.007&0.194&0.008&0.424&0.004&0.476&0.002  \\
	& SE $(\times 10^{-2})$& 0.133&0.134&0.130&0.131&0.205&0.206&0.415&0.415&0.370&0.371\\
	& ECP (\%)& 83.9&95.7&84.3&95.6&84.7&96.5&86.2&96.0&78.6&96.5 \\
	\multicolumn{12}{c}{} \\
	Kurt & Bias $(\times 10^{-2})$& 0.682&0.032&0.679&0.025&1.685&0.056&0.431&0.014&0.789&0.002 \\
	& SE $(\times 10^{-2})$ &1.576&1.778&1.706&2.103&1.979&2.119&1.197&1.211&0.487&0.492\\
	& ECP (\%)& 89.6&93.7&89.9&94.7&85.2&94.7&94.3&95.4&66.2&95.2  \\
	\multicolumn{12}{c}{} \\
	\multicolumn{12}{c}{$n = 200$, \quad $K = 300$} \\
	Mean & Bias $(\times 10^{-2})$ &0.003&0.003&0.002&0.002&0.001&0.001&0.001&0.001&0.026&0.026  \\
	& SE $(\times 10^{-2})$ &0.219&0.219&0.216&0.216&0.326&0.326&0.771&0.771&0.980&0.980 \\
	& ECP (\%) &94.8&94.8&94.2&94.2&94.2&94.2&94.6&94.6&95.7&95.7 \\
	\multicolumn{12}{c}{} \\
	SD & Bias $(\times 10^{-2})$&0.192&0.008&0.187&0.007&0.288&0.008&0.634&0.002&0.711&0.000\\
	& SE $(\times 10^{-2})$&0.133&0.133&0.130&0.130&0.205&0.205&0.414&0.415&0.370&0.371 \\
	& ECP (\%)& 71.4&95.3&72.6&95.2&72.5&96.2&70.2&96.2&55.7&96.6 \\
	\multicolumn{12}{c}{} \\
	Kurt & Bias $(\times 10^{-2})$&1.035&0.025&1.024&0.012&2.542&0.023&0.631&0.019&1.179&0.008\\
	& SE $(\times 10^{-2})$ &1.511&1.745&1.571&1.984&1.952&2.158&1.203&1.226&0.494&0.502\\
	& ECP (\%)&85.4&93.8&85.0&94.8&75.3&92.6&93.0&96.1&35.6&95.7  \\
	\hline
\end{tabular}
\end{spacing}
\end{table}
\end{landscape}



\end{document}